# Rømer all ears: An auditory analog of Ole Rømer's demonstration of the finite speed of light using smartphones as an acoustic vernier

Laís Florencio de Oliveira[1], Gabriel Antonio Caritá[2], João Teles de Carvalho-Neto[1*]

[1]*Centro de Ciências Agrárias (CCA), Universidade Federal de São Carlos (UFSCar), Rodovia Anhanguera, km 174, 13600-970, Araras, SP, Brazil*

[2]*Instituto de Geociências e Ciências Exatas (IGCE), Universidade Estadual Paulista (UNESP), Av. 24-A, 1515, 13506-900, Rio Claro, SP, Brazil*

[*]Corresponding author: jteles@ufscar.br

November 25, 2026

**Abstract**

We present an undergraduate experiment for teaching Ole Rømer's 1676 method for demonstrating the finite speed of light. Two smartphones emit periodic sound pulses whose synchronization is evaluated directly by the observer's hearing, acting as an "acoustic vernier" that achieves millisecond travel-time resolution without microphones or external sensors. Because the measurement relies strictly on propagation delays between stationary stations, it establishes a direct experimental analogy to Rømer's original argument and dispenses with auxiliary wave phenomena absent from the 1676 analysis, such as acoustic beats or the Doppler effect. The pedagogical strength of the experiment lies precisely in this direct appeal to the senses. Loss of synchronization becomes perceptible starting at a separation of approximately 5 m and grows progressively clearer across several tens of meters. Despite its simplicity, systematic blind measurements determine the speed of sound in air with single-run uncertainties of 2 to 5 m/s (around 1%) and agree within 1.1% with theoretical predictions across ambient temperatures from 9 to 25 °C. Beyond numerical measurement, anchoring the method in explicit sensory perception helps to clarify the recurring conceptual entanglement between Rømer's geometric delay and the kinematic Doppler effect, while highlighting how this same travel-time principle underpins contemporary astrophysical techniques, such as pulsar timing and eclipse timing variations (ETV).

## 1 Introduction

Defending his claim that light travels at a finite speed, Ole Rømer remarked in a September 1677 letter to Christiaan Huygens that *"the motion of light may be considered analogous to the motion of sound."*[1] Establishing that finite speed transformed our comprehension of astronomical scales, space, and time.

Although Aristotle and Descartes regarded light propagation as instantaneous, Galileo Galilei reported the first experimental attempt to time it using shuttered lanterns between distant observers, concluding only that light traveled faster than human reaction time under those conditions.[2] Even so, Galileo unintentionally took the first step toward measuring the speed of light by proposing the eclipses of Jupiter's moons as a celestial clock for determining geographical longitude.[2] Pursuing this goal, the Paris Observatory instituted systematic observations of the Jovian satellites.[3] In August 1676, its leading astronomer, Jean-Dominique Cassini, announced that an inequality in the eclipse times of the Jovian moons might stem from the finite travel time of light—an explanation originating with or developed alongside Rømer—though Cassini soon abandoned the hypothesis.[3–5]

Developing this hypothesis, Rømer demonstrated in late 1676 that a finite speed of light accounted for the systematic delays in the eclipses of Io, the innermost Galilean satellite. For Rømer, these delays could not arise from orbital anomalies at Jupiter because their periodicity followed the synodic cycle of Earth's approach to and recession from Jupiter.[1,6] He successfully predicted that Io's emersion on November 9, 1676, would occur about 10 minutes later than deduced from August observations, and he estimated a maximum delay of 22 minutes (now known to be about 16.6 minutes) for light to cross the diameter of Earth's orbit.[6] Supported by Jean Picard, John Flamsteed, and Huygens,[1,7] Rømer's work in Paris confirmed the finite propagation speed required by Huygens's wave principle, leading the latter to compute, in *Traité de la Lumière*,[8] the first explicit value for the speed of light (about 210,000 to 230,000 km/s in modern units).[3,7]

In the physics teaching literature, Rømer's method is treated predominantly through theoretical and historical discussions,[7,9] often framed as a manifestation of the Doppler effect.[9–11] Direct astronomical replications remain rare in instructional settings, even though recent work marking the 350th anniversary of the discovery demonstrated feasibility by eye with telescopes matching late-seventeenth-century resolution.[12] Such campaigns still demand clear night skies and multi-night tracking, typically spanning months across Jupiter's synodic cycle[13] or at least one to two weeks near quadrature when Io's tabulated period is assumed.[10] Yet as a terrestrial experimental analog, we have identified only the demonstration by Korman, Tobin, and McMahon,[14] who emulated Earth's orbit with a rotating microphone that detects the dynamic acoustic Doppler shift from a stationary speaker.

Turning to sound is pedagogically natural because its propagation speed is nearly six orders of magnitude lower than that of light, bringing the phenomenon to human spatial scales. Historically, determinations of the speed of sound by ear relied either on timing the delay between the flash and report of distant gunfire[15,16]—which in instructional settings is compromised by human reaction time of $\approx 200$ ms[17]—or on timing wall echoes via spoken syllables or synchronized hand claps,[15,18] which demands large open spaces with wide reflecting surfaces. Consequently, undergraduate laboratories usually bypass direct time-of-flight measurements in favor of standing-wave resonance tubes,[19] accessing wavelength rather than the temporal propagation of a pulse. More recently, smartphones have become accessible and versatile tools for physics experiments,[20] but their determinations of the speed of sound still rely on microphone time-of-flight capture[21] or spectral resonance analysis in acoustic tubes.[22]

Here, we take Rømer's analogy at face value and propose an experiment based on directly hearing the delay between periodic sound pulses emitted by two smartphones. One stationary device represents the Io–Jupiter system, while the second accompanies the observer as "Rømer's clock" on Earth, initially synchronized with the first. As the observer walks away, the loss of synchronization is perceived "by ear" starting near 5 m, without microphones or audio-analysis software, reproducing the essence of Rømer's 1676 observation without introducing extraneous wave phenomena such as beats or Doppler shifts. Moreover, blind quantitative trials yield the speed of sound in air with an uncertainty near 1%. Because every reading is taken by an observer at rest, the experiment also provides a concrete setting in which to disentangle Rømer's distance-governed delay from the velocity-governed Doppler effect (Sec. 4), a distinction that remains operative in contemporary astrophysical timing.

## 2 Materials and Methods

### 2.1 Materials and cost

The acoustic sources are standard smartphones running any modern web browser compatible with the *Web Audio API*. Although tested here on standard Android smartphones, the application is platform-independent and uses only the built-in speakers, requiring no external amplification and incurring no cost when students use their own devices. Station distances were measured with a 30 m millimeter-graduated tape measure ($15 to $30 USD), and ambient air temperature was monitored with a portable digital probe thermometer having 0.1 °C resolution and 1 °C nominal accuracy ($10 to $15 USD). For outdoor distances beyond 50 m, we added an optional 3D-printed PLA parabolic acoustic reflector ($3 to $5 USD in material) with 18 cm aperture, 2 cm focal length, and a vertex slot to hold the stationary smartphone.

### 2.2 Experimental setup

The apparatus consists of two smartphones running the *IoTick* application (Sec. 2.3), each continuously emitting a train of identical audio pulses separated by a base period $T$ (Fig. 1(a)).

As shown in Fig. 1(b), one smartphone remains at a fixed position, acting as the remote periodic source analogous to Io's eclipses in the Io–Jupiter system (where each pulse represents a consistent fiducial phase of the eclipse cycle, such as successive emersions, as discussed in Sec. 5), while the second phone is carried by the observer along the path. The position of the observer's ear defines the measurement coordinate $x$ relative to a fixed reference position $x_0$ near the stationary phone, establishing the propagation distance $d = x - x_0$ (with $x_0 = 0$ m, $x$ coincides numerically with $d$). Starting with their ear at $x_0$, the observer shifts the pulse of the handheld phone by an offset $t_0$ until perceiving, by ear, a single fused pulse from both devices. The observer then walks to position $x$, where the finite speed of sound $v$ delays the remote pulses by $\Delta t = d/v$, making the two pulse trains perceptually distinct. Remaining stationary at $x$, the observer can shift the handheld phone's offset to a value $t$ until both emissions merge into a single pulse once again, yielding $\Delta t = t - t_0$. Note that the application has no separate "Earth" or "Io–Jupiter" modes. What distinguishes the devices is that, once emission starts, the stationary phone remains untouched throughout the run, whereas all offset and gain adjustments are performed exclusively on the observer's handheld device.

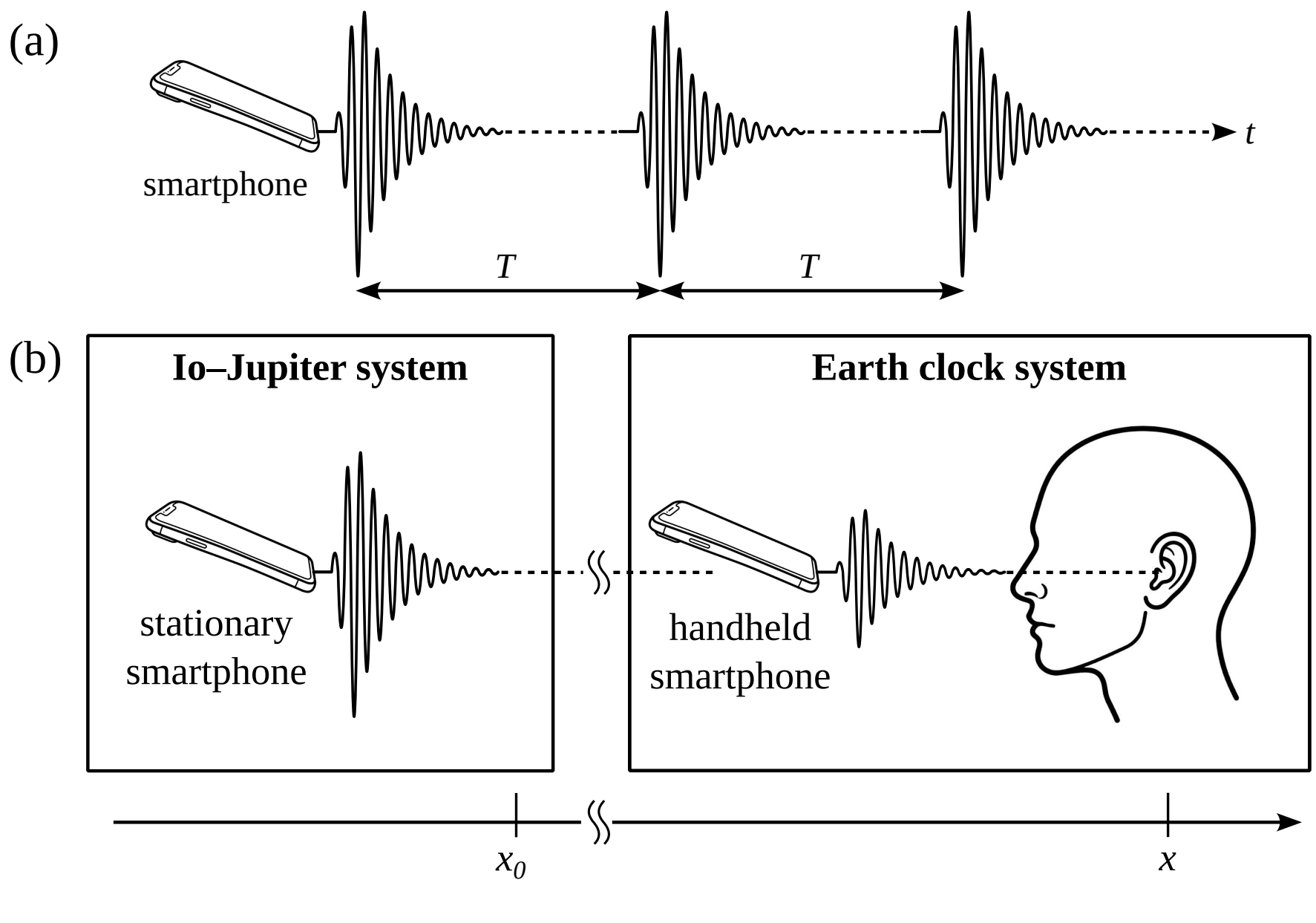

**Figure 1:** (a) Schematic of the periodic audio pulse trains generated by *IoTick* on each smartphone to emulate Rømer's method (although only three pulses are shown, emission is continuous). (b) Experimental configuration in which one smartphone represents the Io–Jupiter system while the second is operated by the observer as the local terrestrial clock. Spatial coordinates correspond to the position of the observer's ear during measurements.

In practice, this adjustment acts as an "acoustic vernier" or, in metrological terms, a temporal coincidence null method. Just as a caliper's vernier replaces the estimation of millimeter fractions with the visual identification of alignment between two marks, *IoTick* reduces travel-time measurement to the auditory identification of coincidence between two pulses. Strictly, a vernier gains resolution from two scales of slightly different pitch, whereas both pulse trains here share the same period, so the analogy concerns the readout alone. Because no reaction to an event is ever timed, the human reaction-time uncertainty that limits manual stopwatch timing[17] drops out entirely, enabling temporal discrimination with millisecond sensitivity.

Three physical and technical aspects govern the success of this implementation. The first is pulse regularity. Over several minutes (the duration of a complete run), spurious clock drifts must remain within a few milliseconds so that the two phones have a stable time base without any inter-device communication. Conventional JavaScript timers suffer tens of milliseconds of jitter under user-interface load. To overcome this, *IoTick* uses the browser's native *Web Audio API* to schedule pulses directly on the device's dedicated audio hardware clock. Combined with offline browser execution, this hardware-level stability usually keeps two autonomous phones synchronized for several minutes. Notably, this independence strengthens the fidelity to the historical episode, which likewise compared two independent periodic phenomena. In 1676, because mechanical pendulum clocks accumulated daily drifts of 10 to 15 seconds,[9] the independent time standard Rømer used to extend the ephemerides from August to November could not be a benchtop clock. Instead, it was Earth's own rotation, determined astronomically from transits of the Sun and stars at the Paris Observatory and interpolated locally by Huygens's pendulum clocks.[3,23] The historical method thus compared, essentially, two autonomous astronomical clocks: the orbit of Io and the rotation of Earth.

The second aspect concerns selecting the acoustic pulse properties that maximize human sensitivity to small arrival-time differences (Sec. 3.1). The third involves spatial range: under moderate background noise, built-in speakers remain audible up to $\approx 50$ m, while mounting the stationary phone at the focus of the parabolic reflector (Fig. 2) partially collimates the sound and extends measurements to $\approx 90$ m (Sec. 3).

### 2.3 The IoTick web application

The periodic acoustic signals are generated by *IoTick*, an open-source Progressive Web Application (PWA) developed for this work, accessible directly through standard mobile browsers and fully operational offline without app-store installation. Rather than acting as an automated sensor or black-box measurement tool, *IoTick* functions strictly as a configurable audio pulse generator and phase-delay controller allowing independent adjustment of the repetition period $T$, carrier frequency $f$, and exponential decay time. During measurements, the observer adjusts a relative time offset $t$ on the mobile device in 1 ms steps cyclically within $0 \leq t < T$ and equalizes perceived loudness via a gain slider. To eliminate confirmation bias during data collection, an optional blind mode (*Hide Records*) conceals numerical readings until data export. Complete interface screenshots, installation protocols, and parameter details are available in the application's repository (see External Repositories).

### 2.4 Uncertainty evaluation and experimental quantities

For every individual reading, we adopt a spatial uncertainty $\delta_x = 20$ cm and a temporal uncertainty $\delta_t = 10$ ms (Type B uncertainty evaluation[24]), which are used for all error bars. Both exceed the 1 mm tape graduation and 1 ms app resolution to account for physiological and postural variability. Specifically, 20 cm reflects the maximum horizontal offset between the observer's foot on the ground marker and their ears, while 10 ms is a conservative estimate of the auditory simultaneity window under field conditions. For a single-distance measurement, the speed and its propagated uncertainty are

$$v = \frac{d}{\Delta t} \quad \text{and} \quad \sigma_v = v\sqrt{\left(\frac{\delta_x}{d}\right)^2 + \left(\frac{\delta_t}{\Delta t}\right)^2} \; . \tag{1}$$

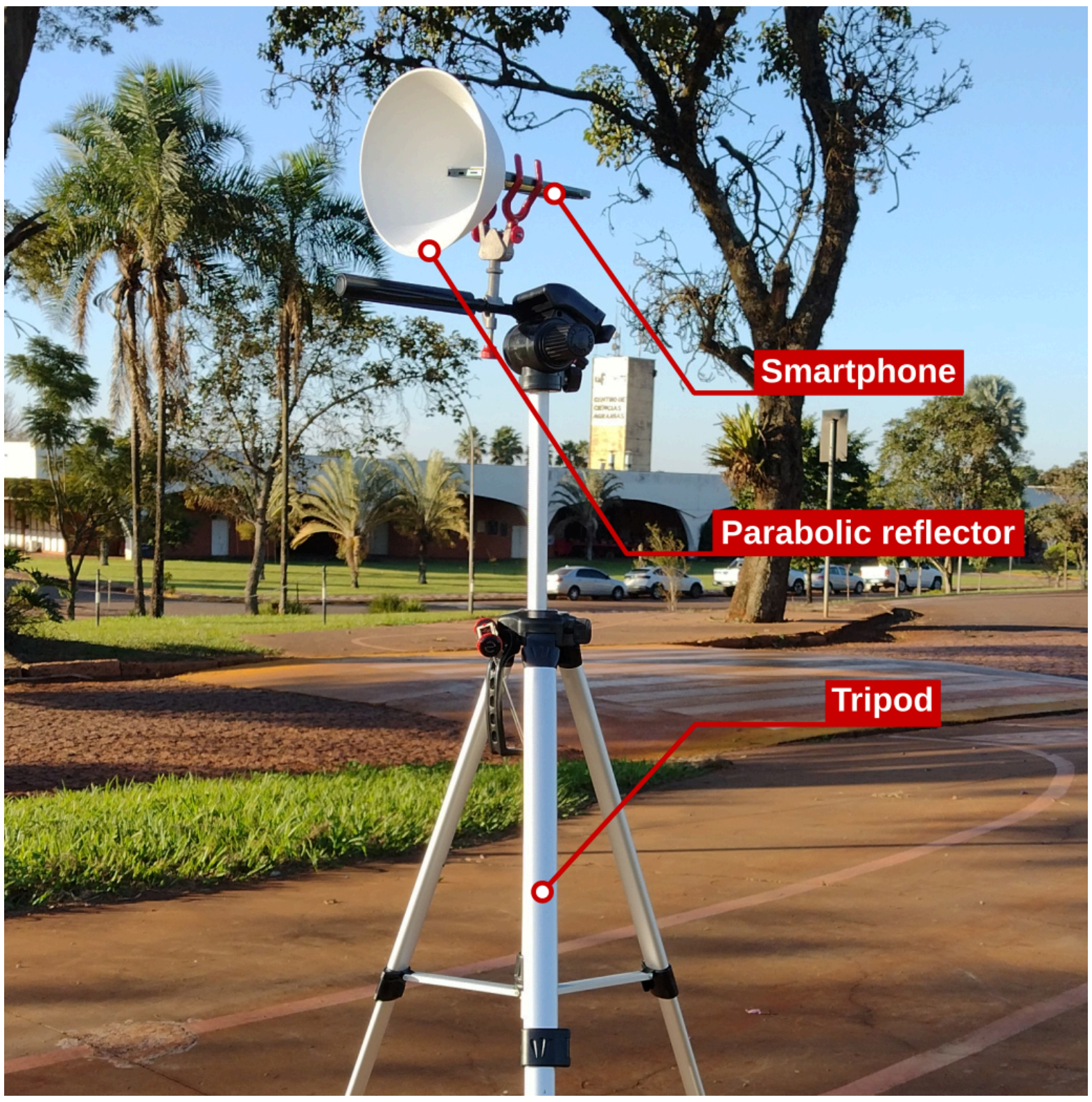

**Figure 2:** Outdoor setup for distances beyond ≈ 50 m, with the main speaker of the stationary Io–Jupiter smartphone positioned at the focus of the 3D-printed parabolic reflector on a tripod. As a convenient mounting solution, a three-prong clamp attached to the tripod handle grips the smartphone horizontally.

Measured sound speeds are compared against the first-order ideal-gas expansion for dry air around 0 °C,

$$v_{\text{th}}(\theta) \approx 331 + 0.6\theta \; , \tag{2}$$

where $\theta$ is the air temperature in °C. Full real-gas models incorporating humidity and barometric pressure[25] differ from Eq. (2) by less than 0.4% at moderate humidity between 8 and 30 °C, well below our experimental uncertainties. We adopt the thermometer's nominal uncertainty of $\delta_\theta = 1$ °C (0.6 m/s in $v_{\text{th}}$), retaining its 0.1 °C resolution only to indicate relative thermal drift within a session, and report speed uncertainties to two significant digits.[24]

# 3 Experimental Results and Discussion

## 3.1 Optimization of acoustic variables and perception of desynchronization

Auditory sensitivity to arrival-time differences depends strongly on the temporal envelope and spectrum of the pulses.[26] In informal exploratory trials among the authors (comparing identical, detuned, and octave sine pairs as well as short noise bursts across 500 to 5000 Hz and 5 to 100 ms decays), distinct pitches or timbres helped distinguish the two phones but substan-tially degraded coincidence acuity. Highest sensitivity was achieved with identical sine pulses at $f = 4000$ Hz, an exponential decay of 50 ms, and a base period of $T = 1.00$ s. The 4000 Hz carrier lies near the peak sensitivity of human hearing[27] and above low-frequency environmental noise from wind and distant traffic. Each pulse begins with a 1 ms attack ramp followed by an exponential decay to $-80$ dB at 50 ms (an effective amplitude time constant of $\tau \approx 5.3$ ms), producing a crisp "tonal click" without broadband artifacts. This 50 ms decay represents an optimal compromise: shorter durations sound excessively faint due to the ear's temporal energy integration,[26] whereas longer durations cause overlapping tails that blur coincidence at short distances.

With these parameters, desynchronization unfolds progressively with distance. For $d < 3$ m, the two pulses fuse perceptually into a single clean tonal click. Around $d \approx 5$ m ($\Delta t \approx 15$ ms), the combined sound loses its sharp onset and acquires a slightly rough or raspy timbre, consistent with auditory sensitivity to fine temporal offsets perceived as timbre changes rather than separate events.[26] As distance increases to between 10 and 15 m ($\Delta t \approx 30$ to 45 ms), two successive clicks resolve clearly, growing to a temporal separation exceeding a quarter of a second at 90 m ($\Delta t \approx 265$ ms).

During field measurements, two practical precautions preserve coincidence acuity. First, because distance attenuation weakens the remote pulse while the handheld phone stays near the ear, the observer must attenuate the local source at each stop via the *Gain* slider (or physical volume buttons) to prevent auditory masking.[27] Second, observers should maintain symmetric hearing geometry by facing the stationary phone directly and holding the mobile device centered in front of the chest, keeping both ears equidistant from both sources.

To allow readers to experience this auditory transition directly, a subtitled benchtop demonstration video (see External Repositories) reproduces field-equivalent delays via *Time Offset*. Figure 3 displays the corresponding *Audacity* audio waveforms across key stages of the demonstration, culminating in blind resynchronization by ear within 3 ms of the initial reference.

### 3.2 Measurements in spatially restricted environments

To test whether sound reflections from walls, floors, and ceilings in typical campus buildings compromise the method, we first carried out a single-distance measurement at $d = 31.0$ m inside a narrow, confined indoor corridor ($\theta = 30$ °C, $v_{\mathrm{th}} = 349.0$ m/s). Auditory nulling gave $\Delta t = 95$ ms, yielding $v = (326 \pm 34)$ m/s via Eq. (1), a 6.6% underestimate consistent with theory within $\approx 0.7\sigma_v$ despite significant reverberation.

A second trial was conducted outdoors along a sidewalk bordering a building facade ($\theta = 29$ °C, $v_{\mathrm{th}} = 348.4$ m/s) with two of the authors operating *IoTick* independently across seven stations from 0.0 to 65.0 m (Fig. 4). Ordinary least-squares fits yielded $v = (321 \pm 24)$ m/s ($R^2 = 0.974$) for Observer 1 and $v = (311 \pm 27)$ m/s ($R^2 = 0.965$) for Observer 2, agreeing with each other within $0.3\sigma$ and with $v_{\mathrm{th}}$ within $1.1\sigma_v$ and $1.4\sigma_v$. This mutual consistency across independent observers, together with the clear linearity ($R^2 > 0.96$), confirms that campus hallways and building walkways remain viable instructional spaces when open fields are unavailable. Nevertheless, all three estimates fall 7% to 11% below $v_{\mathrm{th}}$, a common bias plausibly caused by late-arriving reflections that smear the remote pulse and shift its perceived start toward longer delays.

### 3.3 Systematic blind measurements and temporal consistency test

To test the precision limits of the method away from reflecting walls up to 91.2 m along a tree-lined outdoor walkway using the parabolic reflector (Fig. 2), the authors collected 12 blind series (Table 1) across four periods of a single day spanning 8.1 to 24.7 °C. To eliminate confirmation bias, *Hide Records* remained active throughout, and the nine stopping stations

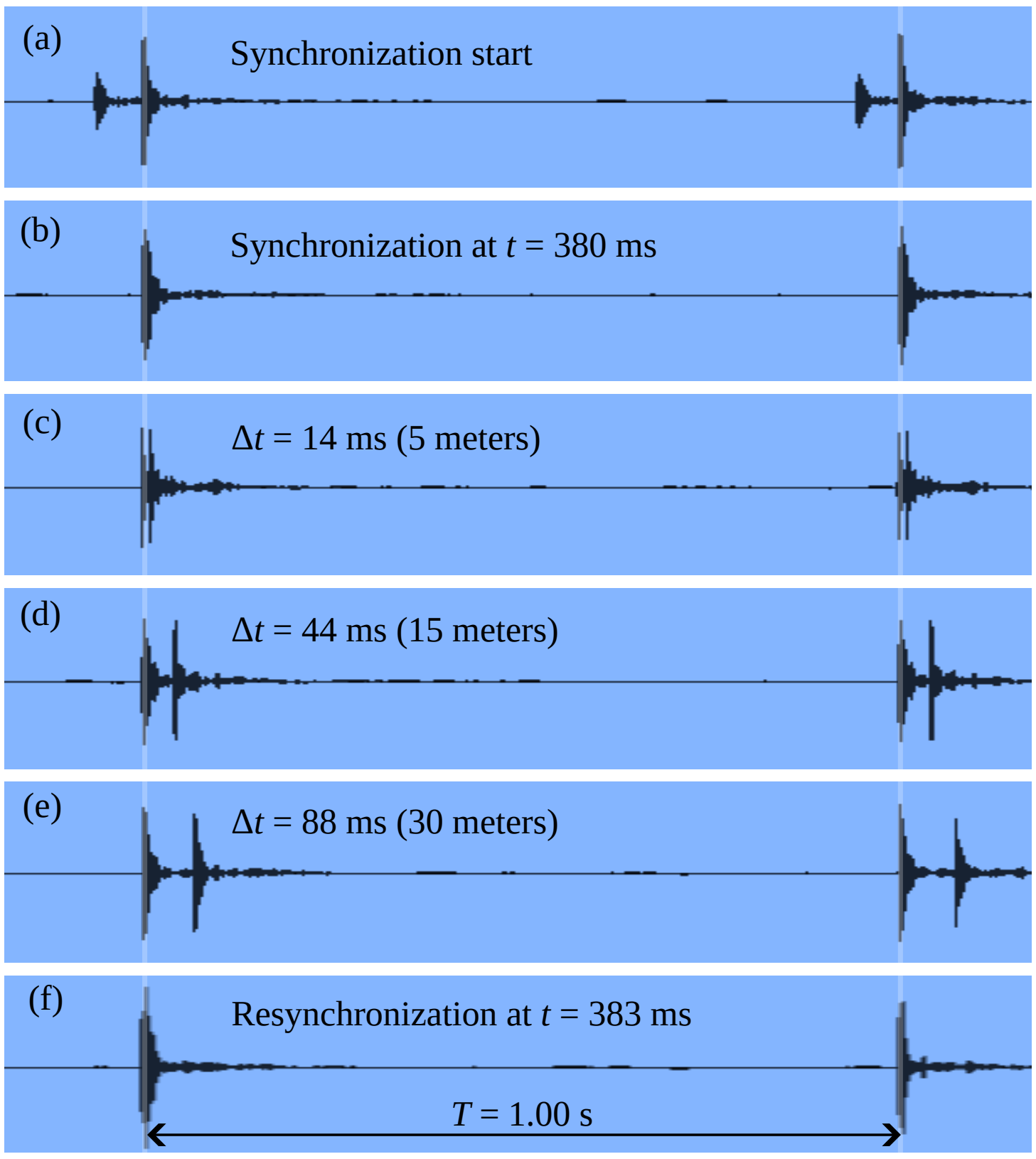

**Figure 3:** Audio waveforms from the benchtop demonstration video showing two consecutive cycles separated by $T = 1.00$ s. (a) Approaching pulses during initial tuning. (b) Initial coincidence achieved at $t_0 = 380$ ms (arbitrary reference). Propagation delays of 14 ms (c), 44 ms (d), and 88 ms (e), corresponding to simulated distances of 5 m, 15 m, and 30 m. (f) Blind resynchronization by ear at $t = 383$ ms ($\Delta t = 3$ ms).

(with $x_0 = 0.0$ m set 0.5 m from the stationary speaker) were marked only visually beforehand. Measuring the path with the tape after all sessions concluded yielded station distances of 0.0, 14.3, 24.4, 38.6, 48.4, 58.6, 69.1, 79.6, and 91.2 m.

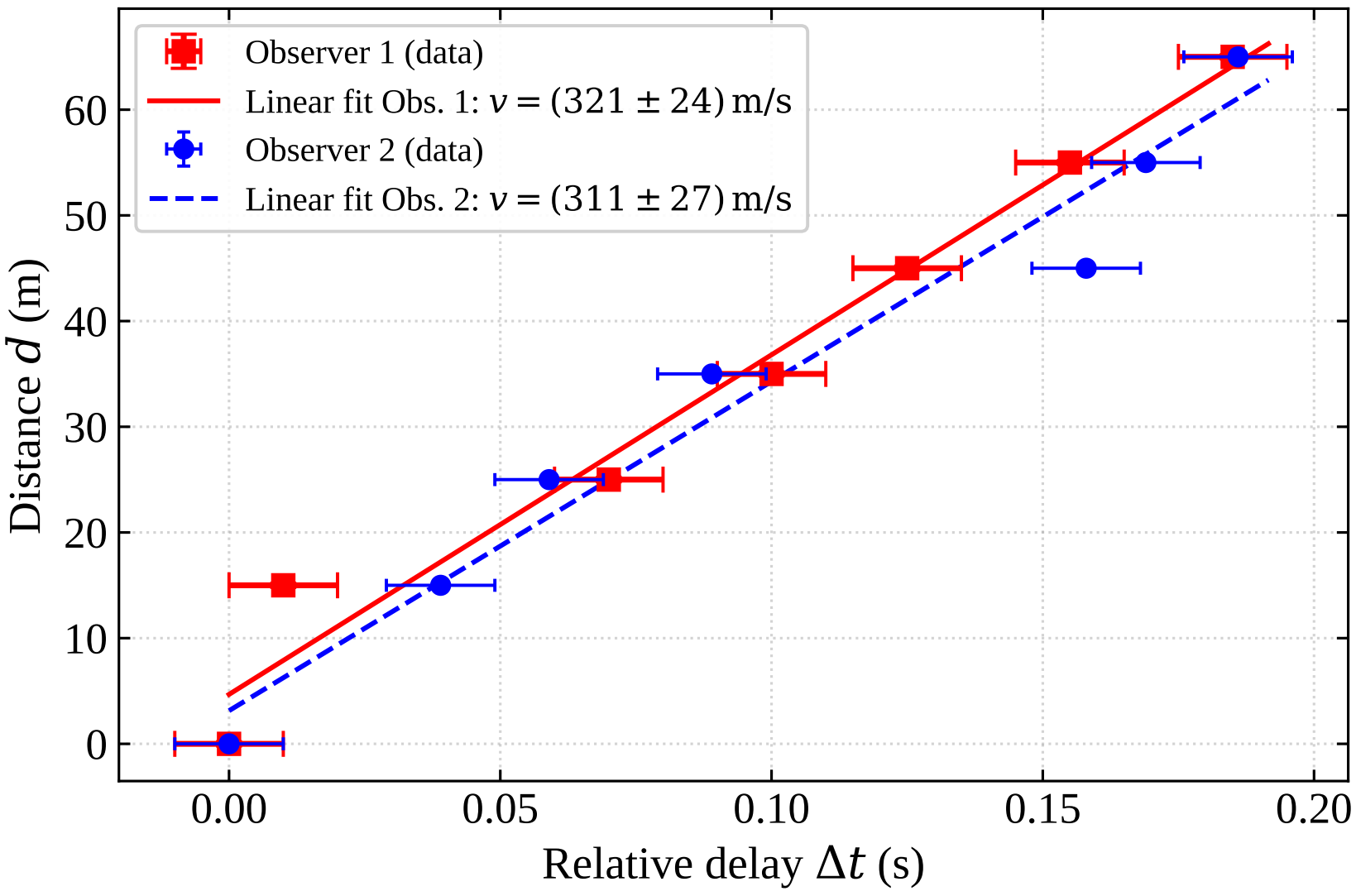

**Figure 4:** Distance as a function of relative time delay for two observers along the exterior facade of a building ($\theta = 29$ °C, $v_{\text{th}} = 348.4$ m/s), with least-squares linear fits.

**Table 1:** Summary of all 12 blind series, where $S$ is the series index, Dur. is the run duration, $\Delta t_{\text{ret}}$ is the return closure error at $x_0$, and $b$ is the linear-fit intercept. Bold rows mark the seven runs satisfying $|\Delta t_{\text{ret}}| \leq 6$ ms analyzed in Table 2 and Fig. 6.

| $S$ | Period | Dur. (min) | $\theta$ (°C) | $\Delta t_{\text{ret}}$ (ms) | $v \pm \sigma_v$ (m/s) | $b$ (m) | $R^2$ |
|---|---|---|---|---|---|---|---|
| 1 | 06:30 | 14 | 8.1–8.3 | +22 | $312 \pm 14$ | 3.3 | 0.9868 |
| 2 | 06:30 | 8 | 8.4–8.6 | −77 | $448 \pm 59$ | 13.4 | 0.8907 |
| **3** | **06:30** | **7** | **9.0–9.5** | **−4** | **$338.1 \pm 4.8$** | **0.7** | **0.9986** |
| **4** | **08:30** | **9** | **13.7–14.2** | **−6** | **$343.3 \pm 4.4$** | **0.2** | **0.9989** |
| 5 | 08:30 | 8 | 14.2–15.2 | −130 | $344 \pm 11$ | 5.8 | 0.9927 |
| 6 | 08:30 | 7 | 15.2–15.4 | −64 | $443 \pm 40$ | −4.6 | 0.9460 |
| **7** | **11:00** | **8** | **20.5–20.8** | **−4** | **$335.9 \pm 4.3$** | **1.0** | **0.9988** |
| **8** | **11:00** | **7** | **20.0–20.8** | **+2** | **$340.3 \pm 4.4$** | **−0.1** | **0.9988** |
| 9 | 11:00 | 7 | 20.0–20.6 | −23 | $362 \pm 12$ | 3.6 | 0.9918 |
| **10** | **11:00** | **6** | **20.6–21.1** | **−2** | **$344.5 \pm 3.7$** | **0.9** | **0.9992** |
| **11** | **16:00** | **7** | **24.7** | **+1** | **$354.3 \pm 2.3$** | **0.5** | **0.9997** |
| **12** | **16:00** | **9** | **23.8–24.7** | **−1** | **$343.3 \pm 3.7$** | **1.8** | **0.9992** |

Because the two phones run autonomously for the 6 to 14 minutes of each series, occasional browser audio interrupts can shift a device's clock reference. We therefore recorded a tenth blind reading upon walking back to $x_0$, defining the return closure error

$$\Delta t_{\text{ret}} = t_{\text{final}}(x_0) - t_{\text{init}}(x_0) \ . \quad (3)$$

Figure 5 illustrates this diagnostic for the three 06:30 series. Whereas Series 3 increased linearly across all nine stations and closed within $\Delta t_{\text{ret}} = -4$ ms, Series 1 and Series 2 suffered mid-walk clock jumps (most visibly near 24.4 m in Series 2) that were diagnosed by their large return deviations (+22 ms and −77 ms).

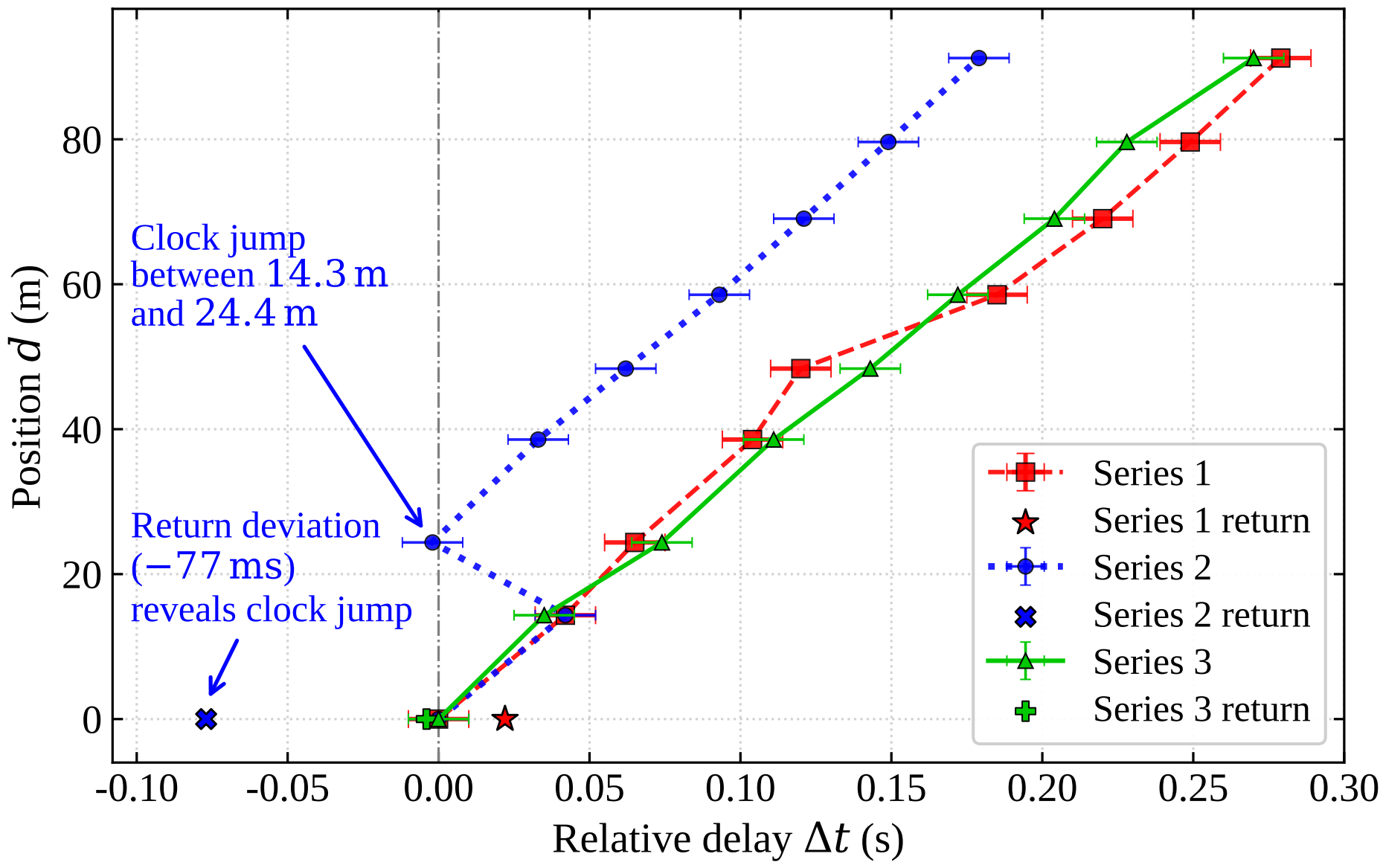

**Figure 5:** Temporal consistency check for the three blind 06:30 series: Series 1 (squares), 2 (circles), and 3 (triangles). Return closure errors are listed in Table 1.

**Table 2:** Comparison between sound speeds from validated blind series ($|\Delta t_{\text{ret}}| \leq 6$ ms) and Eq. (2) at the mean session temperature. Multi-run sessions report the mean and sample standard deviation.

| Period | Series | $\theta$ (°C) | Measured $v$ (m/s) | Theoretical $v_{\text{th}}$ (m/s) | Difference |
|---|---|---|---|---|---|
| 06:30 | 3 | 9.0–9.5 | $338.1 \pm 4.8$ | 336.6 | +0.4% |
| 08:30 | 4 | 13.7–14.2 | $343.3 \pm 4.4$ | 339.4 | +1.1% |
| 11:00 | 7, 8, 10 | 20.0–21.1 | $340.3 \pm 4.3$ | 343.3 | −0.9% |
| 16:00 | 11, 12 | 23.8–24.7 | $348.8 \pm 7.8$ | 345.6 | +0.9% |

Across all 12 series in Table 1, the seven runs satisfying $|\Delta t_{\text{ret}}| \leq 6$ ms yielded linear fits $d(\Delta t) = v\Delta t + b$ with $R^2$ between 0.9986 and 0.9997, speed uncertainties of 2.3 to 4.8 m/s (0.7% to 1.4%), and intercepts $b$ near zero. Because side-by-side coincidence repeatability at $x_0$ is on the order of a few milliseconds, the return check readily detects clock shifts beyond 6 ms. No run fell between 6 and 20 ms, and all five runs with $|\Delta t_{\text{ret}}| > 20$ ms exhibited clear distortions in $v$ or $b$. The 42% rejection rate (5 of 12 runs) highlights the importance of this closure test, as well as the need for preventive measures such as disabling battery-saving modes, keeping screens awake, and closing background apps.

In the multi-run sessions at 11:00 (Series 7, 8, and 10) and 16:00 (Series 11 and 12), even series with high linearity and $|\Delta t_{\text{ret}}| \leq 4$ ms can yield fitted speeds that fall outside their mutual regression uncertainties (such as Series 11 and 12). While the regression uncertainty $\sigma_v$ captures only the internal scatter within a walk, discrepancies between runs likely reflect systematic effects that our protocol cannot isolate, such as a breeze component $u$ along the walkway ($v \pm u$), slow compensating clock drifts during the outward walk, or a small offset between the coincidence judged at $x_0$ and at distant stations. Because $\Delta t = 0$ is fixed at $x_0$, a 4 ms offset of this kind shifts $b$ by about 0.9 m and $v$ by about 3 m/s. Evaluating all validated series in each session via their sample mean and standard deviation (Table 2 and Fig. 6) incorporates this run-to-run dispersion empirically, yielding agreement within 0.4% to 1.1% of Eq. (2) across all temperatures.

## 4 Rømer Versus Doppler: Reflections from Practice

Performing this experiment naturally motivates discussing a recurring question in the history and physics education literature: the relationship between Rømer's method and the Doppler effect. Because the observer stops at each station $x$ to listen and resynchronize the pulses, every measurement occurs at rest relative to the fixed sound source. Moreover, one experiences directly that the delay $\Delta t$ at position $x$ depends solely on the spatial displacement $d = x - x_0$ and not on the speed or pace variations with which the observer walked between stations.

In the literature, however, Rømer's method is frequently associated or equated with the Doppler–Fizeau effect, a link dating back to Tait (1897), who already spoke of the "Döppler-(properly the Römer-) principle."[28] Several authors treat Rømer's method as a historical precursor of the Doppler effect.[9,10,29] This connection is understandable because, in his 1676 text, Rømer used a qualitative argument close to Doppler's principle, noting that during Earth's recession from Jupiter, light takes additional time to reach the observer on each successive revolution, lengthening the apparent interval between consecutive eclipses.[6]

Nevertheless, a sharp physical distinction separates the two effects regarding the quantity actually measured. Historically, Doppler's formulation appeared only in 1842, directly in the wave context, whereas in 1676 the very nature of light was still unsettled and Rømer's delay argument holds for corpuscles and waves alike. Casting Rømer's 1676 analysis as a Doppler measurement therefore invites retrospective anachronism, as even Bailhache, who reconciles Rømer's analysis with an "integrated" Doppler–Fizeau effect, cautions.[29] Observationally, Rømer never resolved single-orbit period shifts. They were a logical deduction from Earth's orbital motion, never measured directly by him. In his 1676 text, he acknowledged that no sensible difference appeared between two single revolutions observed at opposite quadratures, the effect becoming "very considerable" only when many revolutions (e.g., 40) were taken together.[6]

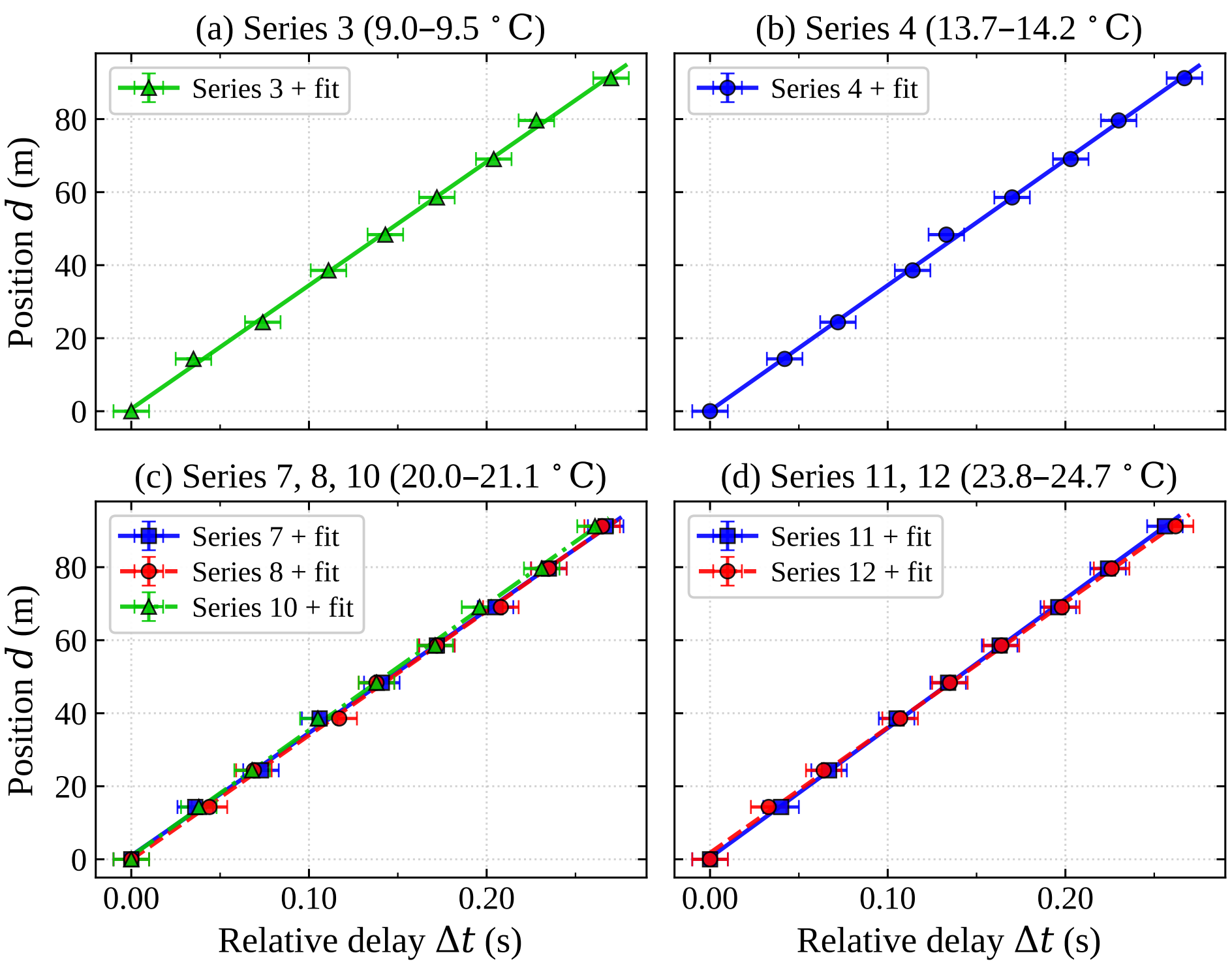

**Figure 6:** Linear fits of position versus relative delay for all seven validated blind series ($|\Delta t_{\text{ret}}| \leq 6$ ms) up to 91.2 m: (a) 06:30, (b) 08:30, (c) 11:00, and (d) 16:00.

Rather than measuring single-orbit shifts, Rømer relied on travel-time delays ($\Delta s/c$) accumulated over dozens of revolutions across a planetary baseline. He compared the November 9, 1676 emersion against ephemeris predictions anchored to August observations after opposition.[4,6,23] Even when he contrasted mean periods over multi-orbit groups of immersions and emersions,[1,23] the period shift still came from an accumulated delay divided by the number of intervening orbits, never from single revolutions.

As Schurmann emphasized when objecting to a description of the arrival-time delays of the X-ray pulsar Hercules X-1 as a "Doppler shift," terminological precision is essential here.[30] The Doppler effect relates the emitted frequency $f_0$ to the detected frequency $f$ through the instantaneous velocities of source and detector, reducing to $f = f_0$ whenever they are at rest relative to each other. Rømer's effect, by contrast, is governed directly by signal travel time across space

$$t = t_0 + \frac{s - s_0}{c} \,, \tag{4}$$

where $c$ is the propagation speed ($v$ in our experiment, with $s - s_0 = d$), $s_0$ and $s$ are the reference and new source–observer distances, and $t_0$ is the predicted arrival time of the $N$th subsequent eclipse (or pulse) had the distance remained $s_0$. The two effects are related as an integral and its integrand. The fractional period shift of an observer with radial velocity $v_r$, $\Delta T/T \approx v_r/c$, is the rate at which this delay accumulates, and summing it over $N$ revolutions returns $(s - s_0)/c$ whatever the velocity history.[29] Rømer and our observer measure this endpoint quantity, not its integrand. Even modern telescopic replications framed via the Doppler effect[10] must accumulate several orbits to overcome eclipse-timing scatter, yielding an expression algebraically equivalent to Eq. (4) for a linearized orbital displacement.

In seventeenth-century astronomy, Keplerian orbital kinematics simply provided the most accurate coordinate map to compute the change in interplanetary distance $(s - s_0)$ over several months, acting as a spatial transport mechanism that placed the observer at distinct locations. In principle, those coordinates could be determined by purely static geometry, much like the parallax measurements used by Cassini and Flamsteed in the 1670s to scale the solar system,[31] without measuring or integrating velocities. If two observers remained stationary relative to Jupiter at Earth's August and November 1676 coordinates—analogous to two static observers at $x_0$ and $x$ holding smartphones in our experiment—the arrival-time difference of the same emersion between the two points would be strictly identical to the delay Rømer observed. Measuring Rømer's effect is therefore distinct from measuring the Doppler effect, just as determining the distance between two railway stations does not depend on the speed of a locomotive traveling between them.

In contemporary astrophysics, this distinction remains sharp in techniques where physical or observational constraints preclude radial-velocity spectroscopy, yet time-of-flight variations of periodic signals can be precisely measured.[32] In the astrophysical literature, the geometric delay $\Delta s/c$ from a source's displacement along the line of sight is literally termed the "Rømer delay" or the "light-travel time effect" (LTTE).[32,33] Here, the primary quantity measured is not the continuous integral of a Doppler radial-velocity curve, but the discrete difference between the observed arrival time of a signal and an unperturbed ephemeris. Its most direct application occurs in pulsar timing, the technique responsible for discovering the first confirmed exoplanets, orbiting the millisecond pulsar PSR B1257+12.[34] Because a neutron star lacks the narrow optical spectral lines required for conventional Doppler shifts, planet detection relied exclusively on periodic modulations in the arrival times of radio pulses as the pulsar orbits the system barycenter. This is a direct application of Eq. (4) in which, once Earth's orbital motion is subtracted, the periodic emitter rather than the observer displaces along the line of sight.

Although the reflex motion of a pulsar binary or an eclipsing binary orbited by a third body (monitored via eclipse timing variations, ETV)[33] may still invite analogies with a discrete-pulse Doppler effect through the periodic compression and expansion of the pulse or eclipse cadence,[35] not even that analogy holds for the internal Rømer delay measured within an isolated, circular-orbit eclipsing binary composed of two stars of unequal masses and radii.[36,37] Because the two components orbit their common barycenter at different radii $(a_1 \neq a_2)$, the two stars exchange places along the line of sight. At the primary eclipse $E_1$, the larger star lies $a_1$ behind the barycenter and the smaller one $a_2$ in front of it. The reverse occurs at the secondary eclipse $E_2$. Neither event train is compressed or expanded, since both repeat with the same orbital frequency $1/P$. Instead, the difference in light-travel path shifts $E_2$ from the midpoint $P/2$ by a constant Rømer delay $\delta t = 2(a_2 - a_1)/c$, as observationally confirmed ($|\delta t| \approx 4.2$ s) in the double white-dwarf binary NLTT 11748.[38] This mirrors the signal structure seen by our stationary observer, who compares two unshifted 1 Hz pulse trains separated by a fixed geometric delay. Moreover, in a circular orbit both eclipses occur at conjunction, where the motion is purely transverse and radial velocities vanish ($v_r = 0$). Even more strikingly, in a hypothetical binary of white dwarfs with featureless continuous spectra (spectral type DC), spectroscopic Doppler measurements would be impossible anywhere along the orbit, whereas photometric eclipse timing would still yield the Rømer delay.

From a pedagogical perspective, linking Earth's orbital dynamics, Io's eclipses, and the Doppler effect has genuine value at an advanced stage of study, particularly when the goal is to deepen understanding of the Doppler effect once students have mastered the underlying kinematic and wave concepts. In that context, Viennot and Leroy-Bury reported that third-year physics undergraduates and trainee teachers valued their graphical space–time linkage once they overcame the initial difficulty of interpreting the graphs.[11] At the introductory level, however, direct empirical studies of joint Rømer–Doppler instruction remain scarce, and many prospective secondary teachers in the same study questioned its suitability for younger students a priori.[11] Established findings in physics education research reinforce this caution. Introductory students persistently confound position with velocity,[39] and in their own survey of third-year university students, Viennot and Leroy-Bury found that only half recognized that the Doppler shift is independent of distance.[11] When teaching Rømer's method at this level, introducing a velocity-governed frequency shift (proportional to $v_r/c$) alongside a distance-governed delay ($\Delta s/c$) can therefore blur both concepts, add extraneous cognitive load through unnecessary interacting elements,[40] and invite misleading analogical transfer from preexisting Doppler misconceptions.[41] Treating the travel-time delay on its own first, while reserving the Doppler synthesis for a later stage, thus preserves both pedagogical clarity and historical fidelity to Rømer's 1676 geometric insight.

## 5 Suggested Instructional Approaches

We suggest two complementary teaching approaches grounded in our apparatus trials across both spatially restricted campus environments and open outdoor baselines. Neither approach has yet been tested in a formal classroom setting, and evaluating student usability and conceptual learning gains remains for future work. The first approach is a qualitative and historical reproduction of Rømer's method in a 15 to 30 m hallway or courtyard that fits within a single class period. Students synchronize the phones side by side (analogous to Rømer's August 1676 calibration near opposition), hear the two pulses separate increasingly with distance (analogous to the November 1676 delay), and hear them merge again back at the starting point. In the astronomical setting, Jupiter's opaque body prevents observers on Earth from seeing both edges of the same eclipse. Only emersions are visible as Earth recedes from Jupiter toward conjunction and only immersions as it approaches opposition,[4,13] so Rømer compared emersions strictly within the receding branch in late 1676.[6] In the acoustic analog, the stationary phone instead maintains a continuous pulse train on both the outward and return walks (analogous to Earth's recession and approach), abstracting away the shadow crossing so that students focus on signal propagation time.

Guiding questions make the historical parallel explicit. (i) *Which of two successive pulses comes from the stationary phone, and what does each device represent?* Only the remote pulse travels $d$, so it is the later one, as students can confirm by briefly lowering the handheld gain. The stationary phone plays Io's eclipses, whereas the handheld phone stands not for a signal but for Rømer's terrestrial clock and the eclipse times he predicted with it. (ii) *What would students hear if sound propagated instantaneously, as Aristotle and Descartes held for light, and what would Rømer have seen?* The pulses would stay aligned at every distance, just as the eclipses would have kept to their predicted schedule. (iii) *Would synchronizing at $x_0 = 15$ m change the offset heard at the same $d$?* No, because the synchronization already absorbs the delay $x_0/v$, leaving only the change in distance audible. Likewise, had Rømer hypothetically timed eclipses of Saturn's moon Titan, discovered by Huygens in 1655, light would take twice as long to reach Earth at opposition (about 70 instead of 35 min), yet the delays accumulated from opposition, set only by the change in Earth–Saturn distance, would be nearly the same. (iv) *If the pulses failed to merge again at $x_0$, what would that indicate?* It would point to a clock shift or an operator error, since a propagation delay depends only on position and vanishes on return, which is also the basis of the closure check in the second approach. By the same logic,

Rømer ruled out orbital anomalies at Jupiter because the delays followed Earth's approach to and recession from Jupiter. (v) *Would the offset change if students walked along a circle centered on the stationary phone?* No, because the distance stays fixed. Similarly, Earth sweeps a long orbital arc between opposition and conjunction, yet only the resulting 2 AU change in Earth–Jupiter distance delays the eclipses. Where time and space allow, students can then test their answers experimentally, except for the counterfactual question (ii).

The second approach turns the experiment into a quantitative outdoor laboratory over 30 to 50 m (or up to 90 m with the parabolic reflector) using the blind mode (*Hide Records*). Students collect several series, match the local gain at each stop, validate runs with the return closure check (Eq. (3), $|\Delta t_{\mathrm{ret}}| \leq 6$ ms), and compare their linear fits with Eq. (2), engaging directly with confirmation-bias control, uncertainty propagation, and systematic error. Depending on the instructor's goals, this activity can be extended, for example, by repeating series at different times of day to probe the temperature dependence of $v$ or by comparing the result with a standing-wave resonance-tube measurement.

Effective implementation of these approaches nonetheless faces at least three practical limitations. First, quantitative precision near 1% requires baselines of several tens of meters, moderate background noise, and calm, dry weather for outdoor runs. Although restricted-space trials confirm that open fields are unnecessary (Sec. 3.2), reverberant hallways introduce modest systematic biases. In larger classes, acoustic cross-talk can be avoided by distributing teams across separate corridors or outdoor areas. Alternatively, instructors can form larger groups per smartphone pair and rotate turns while other students observe or prepare procedures. Second, operating-system audio interrupts remain a recurring failure mode (42% run rejection in our trials), making battery-setting checks and the return closure test (Eq. (3)) essential. Third, the intrinsic reliance on auditory acuity near 4000 Hz creates a genuine accessibility barrier for deaf or hard-of-hearing students. Alongside collaborative work with hearing peers, future *IoTick* updates could use the handheld microphone to display real-time visual pulse envelopes, making delay and alignment directly observable on screen.

## 6 Conclusion

Three and a half centuries after Rømer wrote to Huygens that the motion of light may be considered analogous to the motion of sound, that analogy can now be heard. Obtaining a consistently linear relationship between distance and the measured time offset that yields the predicted speed of sound serves a dual purpose in this work. Epistemologically, it demonstrates that the physiological perception of synchronized or delayed pulses stems genuinely from the sound wave's travel time through space, rather than perceptual illusions, expectation biases, or auditory subjectivity. This agreement thereby legitimizes the pedagogical analogy to Rømer's 1676 discovery, assuring students that the perceived loss of synchronization objectively reflects the finite propagation speed of the signal. Methodologically, this same experimental consistency qualifies the acoustic vernier as a reliable and accurate alternative for measuring the speed of sound in undergraduate physics instruction, transforming what is often taught as an abstract historical milestone into an accessible, human-scale physical measurement.

## External Repositories

The open-source *IoTick* Progressive Web App and its complete source code are hosted at https://github.com/jocoteles/IoTick and permanently archived under https://doi.org/10.5281/zenodo.24468313. In addition, the subtitled benchtop synchronization demonstration video is permanently archived and directly playable at https://doi.org/10.5281/zenodo.24468802.

## Acknowledgments

This work was supported by the Conselho Nacional de Desenvolvimento Científico e Tecnológico (CNPq, Brazil) under Grant No. 407086/2022-6.

Generative AI was used to assist with English language polishing and grammatical accuracy. The scientific content is entirely the work of the authors.